 \documentclass[preprint,aps,prd,nofootinbib,eqsecnum,superscriptaddress,
 preprintnumbers,showpacs]{revtex4}
\usepackage{epsfig}

\begin{document}

\preprint{CU-TP-1178, KIAS-P07025}

\title{Thermal derivation of the Coleman-De Luccia tunneling prescription}
\author{Adam R. Brown}
\email{arb2115@columbia.edu}
\affiliation{Physics Department, Columbia University, New York, New
York 10027, USA}
\author{Erick J. Weinberg}
\email{ejw@phys.columbia.edu}
\affiliation{Physics Department, Columbia University, New York, New
York 10027, USA}
\affiliation{School of Physics, Korea Institute for Advanced Study,
207-43, Cheongnyangni2-dong, Dongdaemun-gu, Seoul 130-722,
Korea}

\pacs{04.62.+v, 11.10.-z, 11.15.Kc, 98.80.Cq}

\begin{abstract}

We derive the rate for transitions between de Sitter vacua by treating
the field theory on the static patch as a thermal system.  This
reproduces the Coleman-De Luccia formalism for calculating the rate,
but leads to a modified interpretation of the bounce solution and a
different prediction for the evolution of the system after tunneling.
The bounce is seen to correspond to a sequence of configurations
interpolating between initial and final configurations on either side
of the tunneling barrier, all of which are restricted to the static
patch.  The final configuration, which gives the initial data on the
static patch for evolution after tunneling, is obtained from one half
of a slice through the center of the bounce, while the other half
gives the configuration before tunneling.   The formalism makes no
statement about the fields beyond the horizon.

This approach resolves several puzzling aspects and interpretational
issues concerning the Coleman-De Luccia and Hawking-Moss bounces.
We work in the limit where the back reaction of matter on metric
can be ignored, but argue that the qualitative aspects remain
in the more general case.  The extension to tunneling between
anti-de Sitter vacua is discussed.

\end{abstract}

\maketitle

\section{Introduction}
\label{intro}

Although it has long been a subject of considerable interest, the
problem of transitions between field theory vacua in curved spacetime
has received renewed attention in recent years, inspired in part by
the possibility of a string landscape containing a vast number of
metastable vacua.

The problem was first addressed by Coleman and De Luccia
(CDL)~\cite{Coleman:1980aw}, who generalized the flat space Euclidean
bounce formalism~\cite{Coleman:1977py,Callan:1977pt} for calculating
the rate at which true vacuum bubbles nucleate within a false vacuum.
When the relevant mass scales are much smaller than the Planck mass
and the flat spacetime bubble size is small compared to the spacetime
curvature, the CDL formalism gives a small gravitational correction to
the bubble nucleation rate, as might be expected.
However, as
gravitational effects become larger, some unexpected features appear
that give one pause.  If the initial false vacuum has positive vacuum
energy, then the matter field within the bounce is nowhere equal to
its false vacuum value, and the bounce itself is completely
insensitive to the shape of the scalar field potential near the false
vacuum.  Further, the formalism appears to involve the fields in
regions beyond the de Sitter horizon and to describe nucleation
processes that take place on a complete spacelike slice of de Sitter
spacetime, even though it is generally understood that this
``somehow'' can't quite be so.  In addition, one finds a greater
variety of bounce solutions than in the flat spacetime case, including
both the homogeneous Hawking-Moss (HM) solution~\cite{Hawking:1981fz}
and families of ``oscillating
bounces''~\cite{Banks:2002nm,Hackworth:2004xb}; yet, 
there are also examples of theories with no CDL bounce at all.

One can view the puzzles associated with the CDL formalism as being
due to the fact that this formalism was originally proposed by arguing
from analogy with the flat spacetime case, rather than by being explicitly
derived.  As a result, even though the formalism may yield a correct answer, it 
is not quite clear what question it is answering.  It is our goal in this
paper to rectify this situation.  We will do this by formulating a 
well-defined question, and then showing that its solution is given by 
the CDL formalism.  In the course of doing so, we will be led to explanations of
some of the unusual features of the formalism, while other troubling aspects
will be seen to disappear when the solutions are properly interpreted.
 
It should be stressed that this is not simply a matter of putting the
formalism on a firmer footing.  We will find that the generally
accepted scheme for obtaining from the bounce the initial conditions
for the Lorentzian evolution after vacuum
decay~\cite{Coleman:1980aw,Guth:1982pn} is based on a incorrect
understanding of the relation between the Euclidean and Lorentzian
spacetimes and must be modified.  The correct procedure gives these
initial conditions only within the horizon and leads, in some cases,
to rather different results.

\begin{figure}[t]
   \includegraphics[width =8cm]{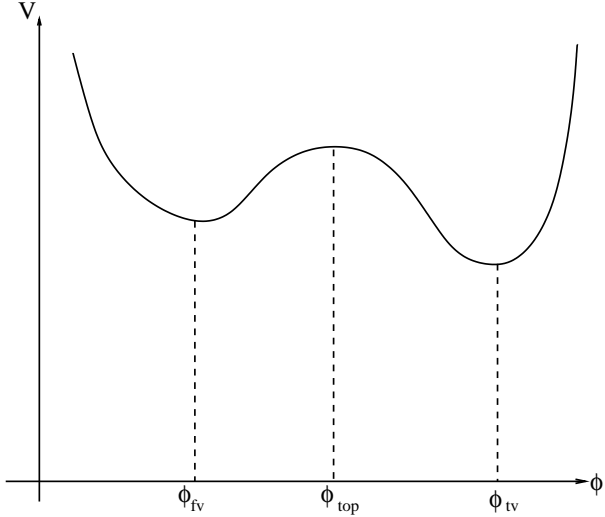}
   \caption{\label{potential} A potential with two minima.}
\end{figure}

Gravity affects vacuum decay in two distinct ways.  First, it requires
that the dynamics of the matter field be worked out in a spacetime
that is curved and that, if the original state is a de Sitter vacuum,
possesses horizons.  Second, the dynamics of the gravitational field
itself must be considered.  In this paper we will, as an initial step,
consider only the former aspect.  To be able to do this in a
self-consistent manner, we consider the theory
of a single scalar field governed by a potential, such as that shown
in Fig.~\ref{potential}, with two unequal minima --- a metastable ``false
vacuum'' and a stable ``true vacuum'' --- but under the assumption
that the variation of the potential in the region between the minima
is small compared to its absolute value; i.e., that
\begin{equation}
     V(\phi_{\rm top}) - V(\phi_{\rm tv}) \ll V(\phi_{\rm tv})  \, .
\label{frozenV}
\end{equation}
With this assumption, we can, to a first approximation, freeze the
metric degrees of freedom and treat the geometry as being a fixed de
Sitter spacetime
with a horizon distance
\begin{equation}
    H^{-1} = \Lambda = \sqrt{3M_{Pl}^2 \over 8\pi V(\phi_{\rm fv}) } \, .
\end{equation}

We will see that many of the troubling aspects of the CDL prescription
are already present in this fixed background limit, and can be
understood without considering dynamical gravity.  The extension of
our methods to the more general case will be discussed in
Sec.~\ref{conc}.  Although we have not yet been able to extend the
technical details of our derivation, we will see that some of the
qualitative features, including the resolution of some
interpretational issues, are readily generalized.

It is possible to describe a portion of de Sitter spacetime, the
causal diamond or static patch, by the time-independent metric\footnote{Our
conventions are such that $ds^2 <0$ for timelike intervals.}
\begin{equation} 
   ds^2 = -\left(1-\frac{r^2}{\Lambda^2} \right) dt^2 
       + \left(1-\frac{r^2}{\Lambda^2} \right)^{-1} dr^2
        +  r^2\left(d\theta^2 + \sin^2\theta\, d\phi^2 \right)  \, .
\label{staticCoord}
\end{equation}
Here $\theta$ and $\phi$ are the usual angular variables on the
two-sphere and $t$ ranges over all real values, but $r$ is restricted
to the range $0 \le r < \Lambda$, with the hypersurface $r=\Lambda$
being the horizon.  With this fixed static background geometry, the
field theory on the static patch is a finite volume system with a
well-defined time-independent Hamiltonian.  However, because the space
is curved, the Hamiltonian density contains position-dependent factors that
would be absent in flat spacetime.  

In addition, the existence of a
horizon leads to a characteristic temperature~\cite{Gibbons:1977mu}
\begin{equation}
    T_{\rm dS} = {1 \over 2\pi \Lambda}  \, .
\end{equation}
To be a bit more precise, if the field on de Sitter space is in the
Bunch-Davies state~\cite{Bunch:1978yq} corresponding to the false
vacuum, then within the static patch one will appear to have a thermal
mixed state with $T=T_{\rm dS}$.  However, the converse need not be
true --- the existence of such a thermal state on the static patch
does not necessarily imply anything about the system beyond the
horizon.

In this paper we will consider vacuum decay within the
framework of this system.  In other words, we will treat the scalar
field on the static patch as a thermal system with $T=T_{\rm dS}$.
The flat spacetime, finite temperature WKB formalism is then readily
adapted to the problem.  This leads to an algorithm for calculating
the rate of vacuum transitions that reproduces the CDL result, but
with some critical changes in interpretation and a new prediction for the 
evolution after tunneling.

A number of previous authors have studied this problem by applying WKB
methods to the calculation of the wave functional of the false
vacuum~\cite{Deruelle:1989qg,Rubakov:1999ir,Gen:1999gi,Fischler:1990pk}.
Our approach differs from these in some important
aspects.  First, we very explicitly restrict our consideration to the
degrees of freedom that lie within the horizon.  The remarkable fact
that the CDL formalism can be recovered from such an approach lends
strong support for our modification of the initial conditions for
post-tunneling evolution.  Second, our approach is based on a thermal
analysis of the curved spacetime field theory.  This leads to a clear
physical interpretation of both the CDL and HM bounces, and the relation 
between the two, and also provides an explanation for some of the most
troubling features of the CDL method.

We begin, in Sec.~\ref{oldstuff}, by reviewing the treatment of vacuum
decay in the absence of gravity, at both zero and nonzero temperature. Next, in Sec.~\ref{cdl}, we apply these methods to the field theory on
the static patch and show how the CDL formalism emerges.  In
Sec.~\ref{variousBounces} we examine a variety of characteristic
bounces in light of this approach.  Next, in Sec.~\ref{evolution}, we
discuss the evolution of the system in the classically allowed regime
after the tunneling process has taken place.  Finally, in Sec.~\ref{conc}, we summarize our results and make some final comments.

\section{Tunneling in flat spacetime}
\label{oldstuff}

The bounce formalism for treating tunneling in the context of a flat
spacetime field theory was developed by Coleman~\cite{Coleman:1977py}
using an approach, based on a multidimensional WKB
approximation~\cite{Banks:1973ps,Banks:1974ij}, that yields the
exponent in the tunneling rate.  The leading subexponential prefactor
terms were obtained by Callan and Coleman~\cite{Callan:1977pt}, using
a path integral approach to the calculation of the imaginary part of
the energy of the false vacuum.  Because finite temperature field
theory can be formulated in terms of a periodic path integral, the path
integral formalism is a natural route for the extension of the
tunneling calculation to finite
temperature~\cite{Langer:1967ax,Linde:1981zj}.  However, because we
are interested in extending these results to curved spacetime, it will
turn out to be more convenient to use a WKB approach to finite
temperature tunneling; we will see that this leads to the same
periodicity requirement on the bounce as the path integral.  The
difficulty with proceeding via the path integral is not so much the
technical issues involved in calculating the determinant factors
(which are ameliorated considerably when working in a fixed background
metric) as the fact that the dilute gas approximation used by Callan
and Coleman cannot be applied when the size of the bounce becomes
comparable to the horizon size.  In addition, the WKB approach has the
added bonus of giving a much clearer physical interpretation of the
bounce.

\subsection{Zero temperature}

Consider first a particle in one dimension with dynamics defined by
the Lagrangian
\begin{equation}
    L =  {1\over 2} M \dot q^2 - U(q)  \, ,
\end{equation}
with $U(q)$ having two unequal minima.  An elementary result in
quantum mechanics is that if the particle has an energy $E$ that is
less than the height of the potential energy barrier separating the
minima, the rate for it to tunnel through that barrier is proportional
to $e^{-B(E)}$, where
\begin{equation}
    B(E)  = 2\int_{q^{(1)}}^{q^{(2)}} dq\, \sqrt{2 M [U(q)-E]} 
\end{equation} 
and
$q^{(1)}$ and $q^{(2)}$ 
are the turning points where $U(q)=E$.

This WKB approximation can be extended to a system with $N>1$ degrees of 
freedom and 
\begin{equation}
    L =  {1\over 2} \sum_{ij} M_{ij} \, \dot q_i \dot q_j  - U(q)  \, ,
\end{equation}
where $U(q)$ again has two minima.  The potential energy barrier separating
the two minima now exists in an $N$-dimensional configuration space,
so we are faced with a multidimensional tunneling problem.  To find
the WKB approximation to the tunneling rate, one considers all
possible paths that start at a point $q^{(1)}_j$ on one side of the
barrier and end at a point $q^{(2)}_j$ on the other side, with the
requirement that $U(q^{(1)}) = U(q^{(2)}) = E$.  For each such path
$P$, one can calculate a tunneling exponent
\begin{equation}
    B(E,q^{(1)},q^{(2)}, P) =
    2 \int_{s_1}^{s_2} ds\, \sqrt{2\Big(U[q(s)]-E \Big)}  \, ,
\end{equation}
where the parameter $s$ along the path is defined so that $ds^2 = \sum_{ij}
M_{ij}dq_i dq_j$ with $q(s_1) = q^{(1)}$ and $q(s_2) = q^{(2)}$.
The leading approximation to the tunneling rate is
obtained from the path and endpoints that minimize $B$.
Taking over a standard result in classical mechanics and inserting a
few sign changes, one readily shows that this minimization
problem is equivalent to finding a solution
of the Euler-Lagrange equations that follow from the Euclidean
Lagrangian            
\begin{equation}
     L_E = {1\over 2} \sum_{ij} M_{ij}\, \dot q_i \dot q_j 
    + U(q)  \, .
\end{equation}
The Euclidean time $\tau$ is simply a reparameterization of the path,
with $q_j(\tau_1) = q^{(1)}_j$ and $q_j(\tau_2) = q^{(2)}_j$.  Because
$dq_j/d\tau =0$ at the end points of the path, the continuation of the
solution beyond $\tau_2$ gives a $\tau$-reversed version of the
original path.  The solution obtained by continuing this back to the
starting point $q_j^{(1)}$ at $\tau_1'$ is known as the ``bounce''.  In particular, 
if $q_j(\tau_1)$ is taken to be the false vacuum minimum itself, one finds
that $\tau_1 = -\infty$, and hence $\tau_1' = \infty$.
Furthermore, the value of $B$ for the optimal path is related to the
Euclidean action 
\begin{equation}
     S_E = \int_{\tau_1}^{\tau_1'} d\tau \, L_E
\end{equation}
of the bounce by 
\begin{equation}
     B(E) =  S_E(q_{\rm bounce}) - (\tau_1'-\tau_1) U(q^{(1)}) =
                   S_E(q_{\rm bounce}) - S_E(q_{\rm init})  \, .
\end{equation}

Finally, we turn to the case of tunneling within the context of a
quantum field theory.  We consider a theory of a single scalar field,
with Lagrangian density
\begin{equation}
   {\cal L} = -{1\over 2} \partial_\mu \phi\, \partial^\mu \phi - V(\phi)
\end{equation}
and $V(\phi)$ of the form shown in Fig.~\ref{potential}.  It is crucial to 
remember that the potential energy is not $V(\phi)$, but rather the
functional 
\begin{equation}
   U[\phi({\bf x})] = \int d^3 x \left[ {1\over 2} 
        ({\bf \nabla}\phi)^2 +V(\phi) \right]  \, .
\end{equation}
The decay of the homogeneous false vacuum proceeds by a tunneling
process in which true vacuum bubbles are nucleated. However, it is not
a tunneling through the barrier in $V(\phi)$ (which would correspond
to a transition from a homogeneous false vacuum to a homogeneous field
configuration on the true vacuum side of the barrier), but rather a
tunneling through the barrier in the infinite-dimensional
configuration space defined by $U[\phi({\bf x})]$, with one end of the
tunneling path being the homogeneous false vacuum and the other end being a
configuration with a bubble of approximate true vacuum embedded in a
false vacuum background.

It is a straightforward process to 
take over the previous results, with the role of the discrete coordinates
$q_j$ now played by the values of the field at each point in space.
For tunneling from an initial configuration $\phi_{\rm init}({\bf x})$, 
with $U[\phi_{\rm init}({\bf x})] = E$, 
the tunneling exponent $B(E)$ is obtained by solving the 
Euclidean field equation
\begin{equation}
    0 = \left({\partial^2 \over \partial \tau^2} 
         +{\bf \nabla^2}\right) \phi - {dV\over d\phi}
\end{equation}
subject to the conditions that 
\begin{eqnarray}
  &&   \phi({\bf x}, \tau_1) =\phi({\bf x}, \tau_1') = \phi_{\rm init}({\bf x})
       \cr \cr
 &&  \left. {\partial \phi ({\bf x}, \tau) \over \partial \tau}\right|_{\tau_1}
     =
    \left. {\partial \phi ({\bf x}, \tau) \over \partial \tau}\right|_{\tau_1'}
     = 0
\end{eqnarray}
for some choice of $\tau_1$ and $\tau_1'$.
A slice through the middle of the bounce, along the hypersurface
$\tau_2 =(\tau_1' + \tau_1)/2$, gives the optimal point for emerging
from the potential energy barrier; i.e., the form of the optimal
bubble.  Thus, $\phi({\bf x}, \tau_2)$ gives the initial conditions
for the classical evolution of the bubble after nucleation.

If the initial configuration is the false vacuum itself, then $\tau_1$
and $\tau_1'$ are $\pm \infty$, and \mbox{$\phi({\bf x}, \tau = \pm \infty)
= \phi_{\rm fv}$}.  Because the intermediate configurations must have
the same potential energy as the initial configuration, the field at
spatial infinity must also approach the false vacuum; i.e.,
$\phi(|{\bf x}|=\infty, \tau) = \phi_{\rm fv}$.  The tunneling
exponent is then
\begin{equation}
    B = S_E(\phi_{\rm bounce}) - S_E(\phi_{\rm fv})  \, .
\end{equation}

\subsection{Nonzero temperature}

Let us now turn to the case of nonzero temperature, beginning again with 
a particle with one degree of freedom.  We will assume that the temperature
$T$ is much less than the height of the potential energy barrier, so that 
a metastable false vacuum state is possible.  Since this is a thermal 
system, the initial state is a mixed state.

\begin{figure}
   \includegraphics[width =8cm]{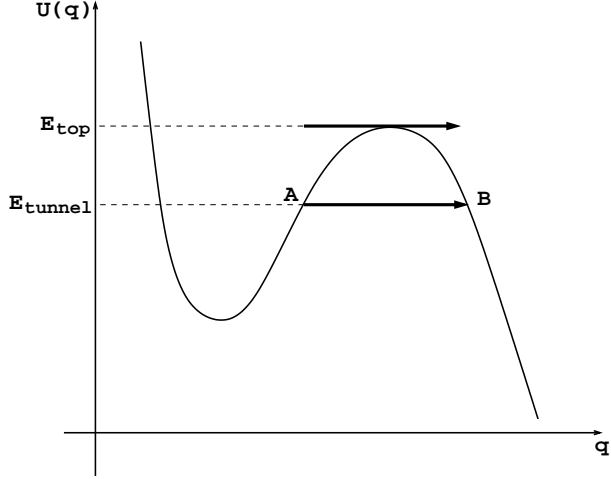}
   \caption{\label{twoModes} A cross-section through 
the potential energy barrier, illustrating the 
two modes of escaping from the false vacuum: thermal
excitation to A, followed by tunneling to B; and thermal 
excitation to the top of the barrier.}
\end{figure}

In this thermal system, the particle need not tunnel from the bottom
of potential well, but can instead tunnel from some thermally excited
higher energy state, as illustrated in Fig.~\ref{twoModes}. The rate
for this thermally assisted tunneling is obtained by taking a thermal
average of the energy-dependent tunneling rates~\cite{Affleck:1980ac}; i.e.,
\begin{equation}
    \Gamma_{\rm tunn} \sim  \int_{E_{\rm fv}}^{E_{\rm top}} dE
         \, e^{-\beta (E -E_{\rm fv})} e^{-B(E)}
     \sim e^{-\beta (E_* -E_{\rm fv})} e^{-B(E_*)}   \, ,
\end{equation}
where $E_*$ is the value of the energy that maximizes the integrand; i.e.,
the value for which $\beta (E -E_{\rm fv}) + B(E)$ is a minimum.

More generally, with many degrees of freedom, we must minimize
\begin{equation}
      \beta (E-E_{\rm fv}) + 2\int_{s_1}^{s_2}  ds\, 
           \sqrt{2\Big(U[q(s)]-E \Big)}
\label{minimizedTerm}
\end{equation}
with respect to path, endpoints, and energy.  The variations with respect
to the endpoints vanish because the integrand vanishes at both of these. 
From our previous
discussion, we know that minimization with respect to the path 
is achieved by requiring that the path correspond to a
solution of the Euclidean Euler-Lagrange equations.  Thus, 
it only remains to determine the optimal energy by solving
\begin{equation}
      \beta = - 2{d\over d E} \int_{s_1}^{s_2}  ds\, 
   \sqrt{2\Big(U[q(s)]-E \Big)}
\end{equation}
with $q_j(s)$ understood to be along a classical path.  The integral
on the right-hand side depends on the energy through the explicit
appearance of $E$ in the square root and through the implicit energy
dependences of the end points and path.  Neither of the implicit
dependences contribute here: as noted above, the variations with respect to
$q^{(1)}_j$ and $q^{(2)}_j$ vanish because the integrand is zero at
the endpoints; the variation with respect to path vanishes once the path is chosen to be
a classical solution.  We therefore have
\begin{equation}
     \beta = 2\int_{s_1}^{s_2} ds {1\over \sqrt{ 2\Big(U[q(s)]-E
                 \Big)} } 
          = 2\int_{s_1}^{s_2} ds {1\over \sqrt{
             \vphantom{\Biggl(}\displaystyle
                 M_{ij} {dq_i\over d\tau} {dq_j\over d\tau}}}
                        = 2\int_{s_1}^{s_2} ds \, {d\tau
                 \over ds} 
           = 2 (\tau_1 - \tau_2)  \, ,
\label{betaPeriod}
\end{equation}
where the second equality uses the fact that we are working with a
solution of the Euclidean equations.  

Equation~(\ref{betaPeriod}) tells us that the passage through the
barrier must take a Euclidean ``time'' $\Delta \tau = \beta/2$.  We
also know that continuation of the solution past the endpoint gives a
$\tau$-reversed solution back toward the starting point.  With this
continuation, we have a solution that is periodic in $\tau$ with
period $\beta$.  Thus, the prescription for calculating the rate of
thermally assisted tunneling is based on finding a solution of
the Euclidean equations with period $\beta$ that has the additional
property that on at least two $\tau$-slices (which are conveniently
taken to be $\tau=0$ and $\tau=\beta/2$) the $dq_j/d\tau$ all vanish.  
The tunneling exponent is then obtained by integrating the Euclidean action 
for this solution over one full period, so that 
\begin{equation}
    \Gamma_{\rm tunn} \sim  e^{-\left[S_E({\rm bounce})
      - S_E({\rm fv})\right]}  \, .
\label{gammasubtunn}
\end{equation}

However, there is a second possible mode for the transition.   Instead of
being thermally excited part of the way up the barrier and then 
tunneling, the particle can be thermally excited all the way to the
top of the barrier.  Up to pre-exponential factors,
the rate for this process is proportional to the Boltzmann factor,
\begin{equation}
    \Gamma_{\rm therm} \sim e^{-\beta(E_{\rm top} - E_{\rm fv})}  \, .
\end{equation}

When there is more than one degree of freedom, there are many possible
paths over the potential energy barrier.  The lowest of these dominates, with the rate governed by the energy $E_{\rm saddle}$ of the
saddle point on this path.\footnote{ Note that all that is required
here is that the path be a local minimum among paths across the barrier.
There may be higher saddle points that are also relevant because they 
lead to different final states.}
This saddle point is a stationary point of
the potential energy and is thus a time-independent solution of the
ordinary equations of motion or, equivalently, a $\tau$-independent
solution of the Euclidean equations of motion.  Viewing it this way,
we can write
\begin{equation}
      \beta \left( E_{\rm saddle} - E_{\rm fv} \right) = 
          S_E({\rm saddle}) - S_E({\rm fv})    \, ,
\end{equation}
where the actions are understood to be calculated over 
a $\tau$ interval equal to $\beta$.

\begin{figure}
   \includegraphics[height =8cm]{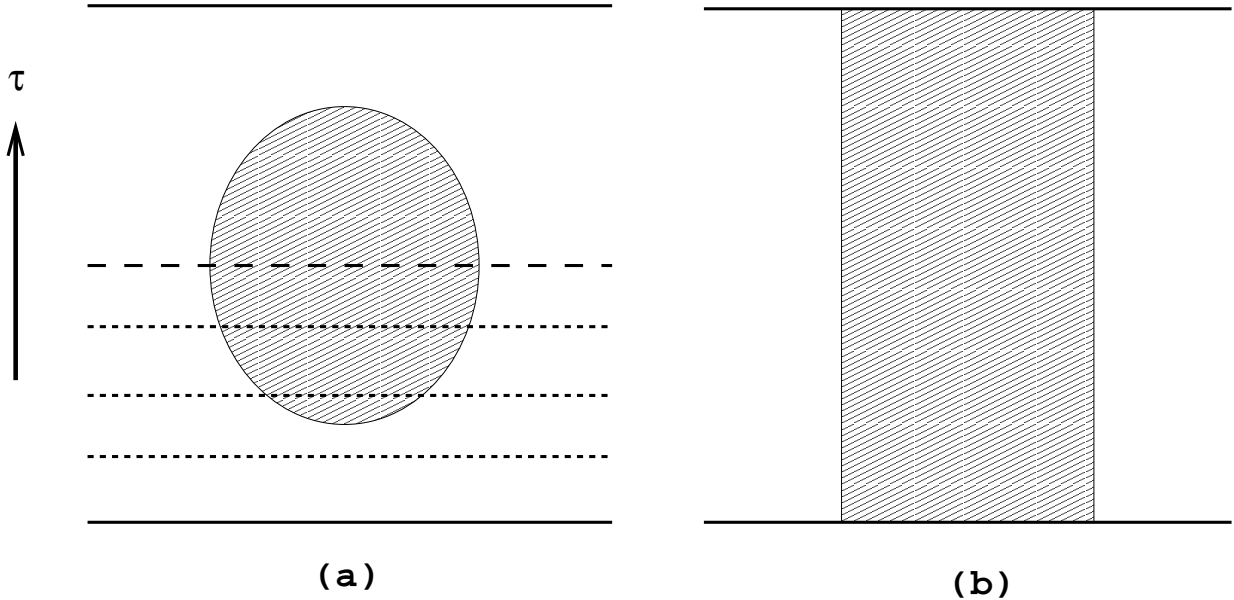}
   \caption{\label{hiTphi} The two types of bounces at finite
temperature for flat spacetime. The shaded areas denote regions where
the field is on the true vacuum side of the barrier.  In both cases
the imaginary time $\tau$ runs vertically, while the horizontal
direction represents the three spatial directions.  In each diagram
the top and bottom solid lines are identified, making $\tau$
compact. The bounce in (a) corresponds to thermally assisted tunneling
from the approximately false vacuum configuration on the $\tau$ slice
represented by the solid lines to the configuration on the $\tau$
slice indicated by the dashed line.  These two configurations are
connected by a series of intermediate configurations, corresponding to
the dotted lines.  The $\tau$-independent bounce in (b) corresponds to
thermal excitation over the barrier.  A constant-$\tau$ slice through
this bounce gives a critical bubble configuration, which is a saddle
point on the potential energy barrier.}
\end{figure}

Since a $\tau$-independent solution can be viewed as being periodic
with any period, the prescription to seek a Euclidean solution with
period $\beta$ actually covers both transition modes, with the
dominant mode being determined by the value of the Euclidean
action. Figure~\ref{hiTphi} illustrates the two types of solutions in
the field theory setting, for the case where $\beta$ is somewhat
larger than the characteristic size of the zero-temperature bounce.
The bounce for thermally assisted tunneling, shown in
Fig.~\ref{hiTphi}a, is a somewhat deformed version of the
zero-temperature bounce.  Note in particular that, in contrast with
the zero-temperature case, the initial configuration, given by the
$\tau=0$ hypersurface, is not identically equal to $\phi_{\rm fv}$.
Figure~\ref{hiTphi}b shows the saddle point solution; a slice along
any hypersurface of constant $\tau$ gives the configuration of a
critical bubble.

It can be shown~\cite{Coleman:1977py} that a bounce solution always exists
for the zero-temperature case;
the same arguments can be easily modified
to show that a critical bubble solution of the sort shown in 
Fig.~\ref{hiTphi}b always exists.   However, there is no 
guarantee that the $\tau$-dependent bounce of Fig.~\ref{hiTphi}a
will persist for all values of $\beta$.  The absence of 
such a bounce would correspond to a situation in which the 
expression in Eq.~(\ref{minimizedTerm}) had no minimum, but instead
was monotonically decreasing as $E$ 
varied from $E_{\rm fv}$ to $E_{\rm saddle}$.

\section{Thermal tunneling on the static patch}
\label{cdl}

We now turn to the field theory on the static patch of de Sitter spacetime.
In terms of the coordinates and metric of Eq.~(\ref{staticCoord}), the action for our scalar
field theory takes the form
\begin{equation}
     S = \int d^4x \sqrt{-\det g} \left[ -{1\over 2} g^{\mu \nu} \,\partial_\mu\phi
        \, \partial_\nu\phi  - V(\phi) \right]  \, ,
\end{equation}
where the spatial integral is restricted to the region $r < \Lambda$.
If we write the three-dimensional spatial metric as $h_{ij} = g_{ij}$, and define  
\begin{equation}
     - g_{tt} = 1 - {r^2\over \Lambda^2} = A(r)  \, ,
\end{equation}
we can rewrite this action as 
\begin{eqnarray}
    S &=& \int dt \int d^3x \sqrt{\det h} \left[ {1 \over 2\sqrt{A(r)}} 
        \left({d\phi\over dt}\right)^2 
       -{1\over 2}\sqrt{A(r)} \,h^{ij} \,\partial_i\phi\, \partial_j\phi
       -\sqrt{A(r)}\,V(\phi) \right]
     \cr &=& \int dt \, L   \, .
\end{eqnarray}

Putting aside for the moment the origins of this expression in a
curved spacetime, we can choose to view the Lagrangian $L$ as
describing a field theory, on a curved three-dimensional space, whose
interactions happen to have an extra position-dependence arising from
the various factors of $\sqrt{A}$.  The energy functional for this
theory is 
\begin{equation}
   E = \int d^3x \sqrt{\det h} \left[ {1 \over 2\sqrt{A(r)}}
        \left({d\phi\over dt}\right)^2
       + {1\over 2}\sqrt{A(r)} \,h^{ij} \,\partial_i\phi\, \partial_j\phi
       + \sqrt{A(r)}\,V(\phi) \right]
\end{equation}
We can now immediately carry over
the results of the previous section.  In particular, to study vacuum
transitions at $T=T_{\rm dS} =1/(2\pi\Lambda) $, we look for periodic solutions
to the Euler-Lagrange equations of the Euclidean action
\begin{eqnarray}
    S_E = \int_{-\pi\Lambda}^{\pi\Lambda} d\tau \int d^3x \sqrt{\det h} 
            \left[ {1 \over 2\sqrt{A(r)}} \left({d\phi\over d\tau}\right)^2
      + {1\over 2}\sqrt{A(r)} \,h^{ij} \, \partial_i\phi \,
          \partial_j\phi +\sqrt{A(r)}\,V(\phi) \right]  \, .
\end{eqnarray}
Note that, for later convenience, we have chosen the integral over the
periodic variable $\tau$ to run from $-\beta/2 = -\pi\Lambda$ to
$\beta/2 =\pi\Lambda$ rather than from 0 to $\beta$.  We will use the convention that 
the hypersurface $\tau=-\pi\Lambda\sim \tau=\pi\Lambda$ is taken to
give the (approximately false vacuum) configuration before tunneling,
with the hypersurface half a period away, at $\tau = 0$,
giving the configuration after tunneling, and thus the initial
condition for the subsequent classical evolution.

We now restore $A$ to its role as a metric factor, but now as part of a 
Euclidean metric.  Thus, we define 
\begin{eqnarray}
     \tilde g_{ab} dx^a dx^b &=& A \, d\tau^2 + h_{ij}\, dx^i dx^j
      \cr &=& \left(1 - {r^2\over \Lambda^2} \right) d\tau^2 
            + \left(1 - {r^2\over \Lambda^2} \right)^{-1} dr^2 
           + r^2 \left(d\theta^2 +\sin^2\theta\, d\phi^2 \right)  \, .
\label{hopfmetric}
\end{eqnarray}
With $\tau=-\pi\Lambda$ and $\tau =\pi\Lambda$ identified and
$r$ ranging between 0 and $\Lambda$, this is the round metric for a
four-sphere.\footnote{Any other choice of the temperature would have
led to a Euclidean manifold with a conical singularity at $r=\Lambda$.
At least within the framework of our fixed-background approximation,
we see no inconsistency in such a situation; it just happens not to be
the one that is relevant for the thermal state encountered in vacuum
tunneling.}  More explicitly, with the identifications
\begin{eqnarray}
   y^1 &=& r \, \sin\theta \cos\phi  \cr
   y^2 &=& r\, \sin\theta \sin\phi  \cr 
   y^3 &=& r \, \cos\theta    \cr 
   y^4 &=& \sqrt{\Lambda^2-r^2} \, \cos(\tau/\Lambda)  \cr
   y^5 &=& \sqrt{\Lambda^2 - r^2} \, \sin(\tau/\Lambda)
\label{hopfdef}
\end{eqnarray}
this is the metric on a four-sphere of radius $\Lambda$ embedded in five-dimensional
Euclidean space.
Our Euclidean action can now be written as
\begin{equation} 
   S_E = \int d^4 x \sqrt{\det \tilde g} \left[ {1 \over 2}\tilde g^{ab}
     \,  \partial_a \phi
            \,  \partial_b \phi + V(\phi)  \right]
\label{curvedEucAction}
\end{equation} 
and the rate for vacuum decay is 
\begin{equation}
    \Gamma \sim e^{-[S_E({\rm bounce}) - S_E({\rm fv})]}  \, .
\label{curvedrate}
\end{equation} 

Let us compare this with the CDL prescription.  CDL instruct us to
solve the Euclidean equations for coupled matter and gravity.  If
$V(\phi)$ satisfies Eq.~(\ref{frozenV}), then to leading order one can
ignore the effects of the variation of $\phi$ on the metric, which
becomes that of a four-sphere of radius $\Lambda$.  With the
background gravity thus fixed, the equations for $\phi$ are precisely
those following from the action of Eq.~(\ref{curvedEucAction}).

In our thermal tunneling picture we have the additional requirement,
as noted above Eq.~(\ref{gammasubtunn}), that $d\phi/d\tau$ vanish
identically on the hypersurfaces $\tau=0$ and $\tau=
\pi\Lambda$.  In terms of the $y_a$ defined above, the union of these
two hypersurfaces is the three-sphere formed by the intersection of the
$y^4=0$ hyperplane with the four-sphere of radius $\Lambda$.  Although the existence
of such a three-sphere with vanishing $\tau$ derivatives is not
explicitly stated in the CDL prescription, it follows from their
assumption that the bounce has O(4) symmetry.

Once the bounce has been found, CDL determine the tunneling rate from
the combined gravity plus matter Euclidean action.  In the
fixed-background approximation the contributions of the
Einstein-Hilbert term cancel between the bounce and the homogeneous false
vacuum, leaving precisely the result in Eq.~(\ref{curvedrate}).

\begin{figure}[t]
   \includegraphics[height =10cm]{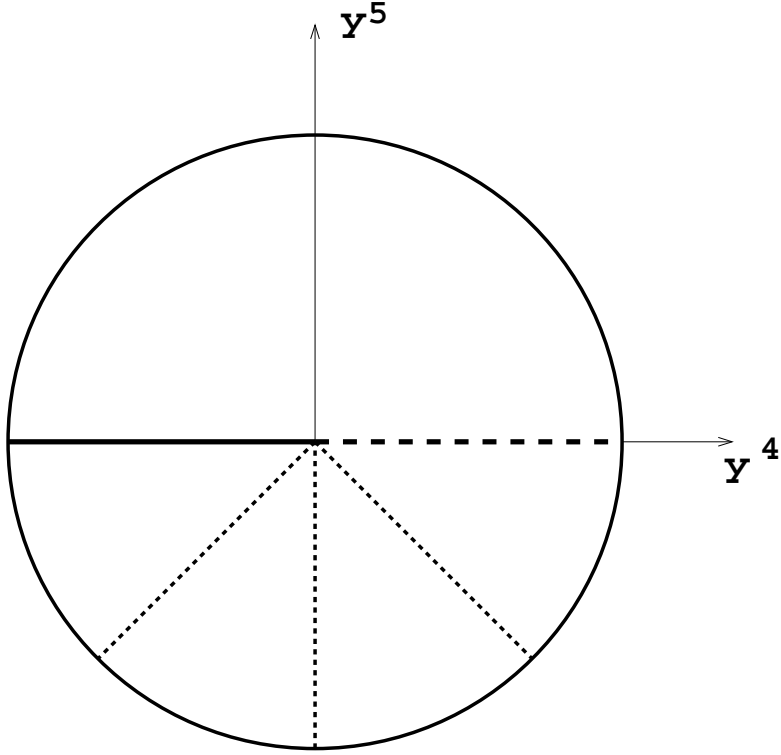}
   \caption{\label{tauSlicing} Slices of constant $\tau$ projected
onto the $y^4$-$y^5$ plane.  The correspondence with those in the
flat-spacetime bounce of Fig.~\ref{hiTphi}a is indicated by the form 
of the lines.  The initial and final configurations of the tunneling
path correspond to the solid and dashed lines, respectively, while
intermediate configurations are obtained from slices along the 
dotted lines. }
\end{figure}

Thus, by treating the field on the static patch as a thermal system,
we have arrived at precisely the CDL prescription for the vacuum decay
rate.  However, our approach leads to a radically different
interpretation of the bounce solution itself.  In our approach,  as illustrated in 
Fig.~\ref{tauSlicing}, the
hypersurfaces of constant $\tau$ provide a foliation of the
four-sphere corresponding to a tunneling path (traversed in both
directions) through configuration space, with the endpoints of the
path given by the hypersurfaces $\tau=0$ and $\tau=\pi\Lambda$.  For
each value of $\tau$, the configuration is specified only on the
region within the horizon; no reference is ever made to quantities
beyond the horizon.  In the CDL description, there is no hypersurface
corresponding to the initial configuration, and the entire hypersurface
bisecting the bounce --- the union of our $\tau=0$ and
$\tau=\pi\Lambda$ hypersurfaces --- is taken to define the final
configuration after emergence from the barrier, and thus the initial
data for the subsequent classical evolution.  This
hypersurface is interpreted as giving initial data on an entire
spatial slice of de Sitter spacetime, including the region outside the
horizon.

We can now also understand what is perhaps the most puzzling aspect of
the CDL formalism, the fact that $\phi$ never achieves its false
vacuum value on the bounce, with the result that the bounce solution is
independent of the shape of the potential in a region near $\phi_{\rm
fv}$.  This is because tunneling from a thermally excited state is
always preferable to tunneling from the false vacuum.  As illustrated
in Fig.~\ref{twoModes}, we can think of this thermally assisted
tunneling as a two-step process in which the field is first thermally
excited to a preferred starting configuration $\phi_A({\bf x})$,
indicated by A in the figure, and then tunnels quantum mechanically
through the barrier to a configuration $\phi_B({\bf x})$. The first
step depends (up to pre-exponential factors) only on the energy
difference between the false vacuum and A, but not on any other
details of the potential energy, including the shape of $V(\phi)$.
The second step depends on the configurations that interpolate between
$\phi_A({\bf x})$ and $\phi_B({\bf x})$; because $\phi$ is nowhere
equal to $\phi_{\rm fv}$ on any of these configurations, the details
of the potential near there never enter.\footnote{For an
illuminating related discussion in the context of a toy model,
see~\cite{Batra:2006rz}.}

\section{A miscellany of bounces}
\label{variousBounces}

It may be helpful to 
examine a variety of characteristic bounces in the light of our
approach.  In all cases we will assume that the bounce has O(4)
symmetry, and in all but the last will take the solution to be
oriented so that this symmetry corresponds to invariance under all
rotations that leave the $y^4$-axis invariant; the solutions will then
only depend on $y^4$.  There are then two distinguished poles, at $y^4=\Lambda$
($r=\tau=0$) and at $y^4=-\Lambda$ ($r=0$, $\tau=\pi \Lambda$).

\begin{enumerate}

\begin{figure}[t]
   \includegraphics[height =8cm]{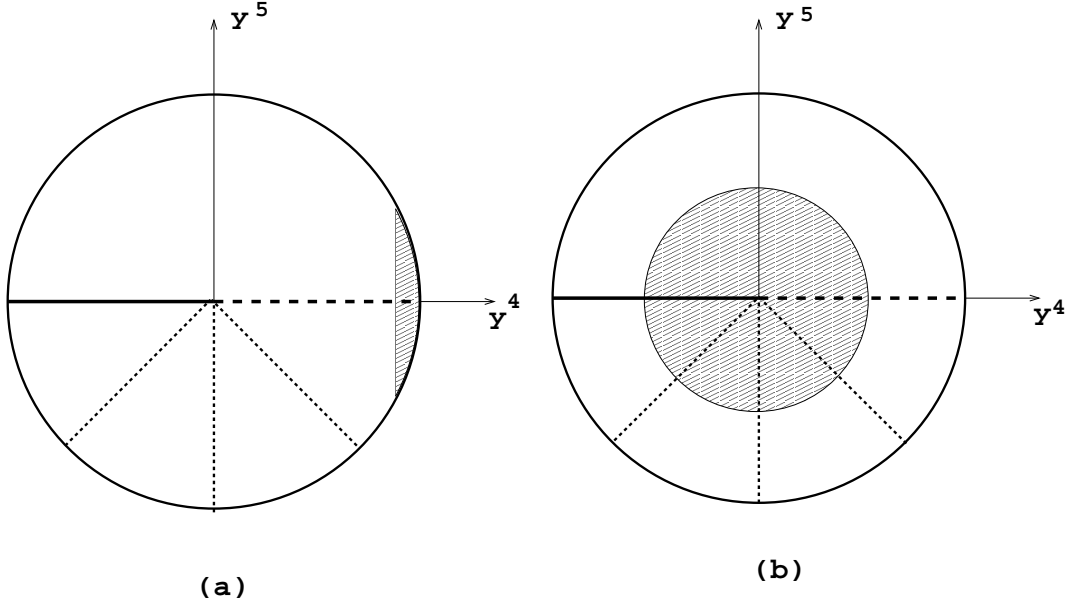}
   \caption{\label{twobounce} Two possible orientations of a CDL
   bounce. In (a), the bounce is centered about the pole at $y^4=\Lambda$,
   and corresponds to nucleation via tunneling of a bubble in the center of 
   the static patch.  In (b), the location of the bounce has been rotated 
   by 90 degrees.  The bounce now corresponds to the purely thermal creation of a true 
   vacuum region that intersects the horizon.    }
\end{figure}

\item{\bf Small CDL bounces:} When the mass scales in $V(\phi)$ are
much less than the Planck mass and the radius of the flat space bounce
is much less than $\Lambda$, the CDL bounce has a region of
approximate true vacuum, centered about the pole at $y^4 = \Lambda$,
that is very similar to the corresponding flat space bounce (see
Fig.~\ref{twobounce}a).  Outside this region $\phi$ rapidly tends
toward --- but never quite reaches --- its false vacuum value, with
$|\phi - \phi_{\rm fv}|$ being smallest at the antipodal point,
$y^4=-\Lambda$.  The initial and final configurations of the tunneling
path, $\phi_A({\bf x})$ and $\phi_B({\bf x})$, are given by the
three-dimensional slices $\tau = \pi\Lambda$ (i.e., $y^4 \le 0$ and $y^5
=0$) and $\tau =0$ ($y^4 \ge 0$ and $y^5 =0$), respectively.
Note that these configurations, as well as all of the interpolating
configurations, overlap at the two-sphere $x^4=x^5=0$, which is just
the horizon, $r=\Lambda$.  The initial configuration differs only very
slightly from the homogeneous false vacuum (with, curiously, the
deviation being greatest at the horizon), while the final
configuration has a true vacuum bubble embedded in a background of
approximate false vacuum.  Apart from the fact that the bounce
corresponds to only a finite region of space, the situation is similar
to that in flat spacetime tunneling.

\item{\bf Large thin-wall bounces:} As the mass scales increase, one
possibility is that the true vacuum region of the bounce occupies a
larger fraction (although always less than half) of the four-sphere,
while the wall separating the true vacuum and false vacuum regions
remains thin.  It was noted some time ago that this bounce can be
interpreted either as mediating decay of the false vacuum via the
nucleation of true vacuum bubbles, or as mediating the creation of a
false vacuum region inside a true vacuum background~\cite{Lee:1987qc}.
We can now sharpen this interpretation.  In the thin-wall limit, we
can assume that away from the wall the field is exponentially close to
one vacuum or the other.  The $\tau=\pi\Lambda$ slice is then a horizon volume
within which $\phi \approx \phi_{\rm fv}$ everywhere, while the
$\tau= 0$ slice is a horizon volume in which there is a true
vacuum bubble surrounded by a false vacuum background.  Taking the
former to be the initial configuration and the latter to be the final
one corresponds to the standard nucleation of a true vacuum bubble.
Reversing the roles of the two slices gives a process in which a
horizon volume of true vacuum is first thermally excited to a
configuration where the region near the horizon is in the false vacuum
and that in the center in the true vacuum, and then tunnels through
the potential barrier to a configuration of homogeneous false vacuum.
Note that the final configuration does not contain a bubble wall.
This does not signify that the false vacuum region covers all of
space, but merely that it fills the horizon volume and that no
statement can be made concerning the interface (beyond the horizon)
between it and other regions.

\item{\bf Thick-wall bounces:} An alternative possibility as the mass
scales increases, one which is typically associated with the case where
$V(\phi)$ is relatively flat near its maximum, is that the wall
separating the two phases becomes wide enough to occupy a significant
fraction of the four-sphere.  This eventually leads to a situation in
which at neither of the poles is the field near a vacuum value.  Let
the values of the field at the poles be $\phi_a$ and $\phi_b$ and its
value at the horizon (and indeed everywhere on the $\tau =
\pi\Lambda/2$ slice) be $\phi_H$, with $\phi_{\rm fv} < \phi_a <
\phi_H < \phi_b < \phi_{\rm tv}$.  Starting from a false vacuum state, say,
the field thermally fluctuates to a configuration with the field equal
$\phi_H$ at the horizon and $\phi_a$ at the center of the horizon
volume, and then tunnels through the potential energy barrier
to a configuration where the field is still $\phi_H$ at the horizon,
but now equal to $\phi_b$ at the center.  As $\phi_a$ and
$\phi_b$ approach the local maximum of $V$ at $\phi_{\rm top}$, the 
transition increasingly becomes predominantly a thermal excitation
process rather than one of quantum tunneling.

\item{\bf Hawking-Moss bounce:} The limiting case of the thick-wall
bounce is the homogeneous HM solution.  Although the original
paper~\cite{Hawking:1981fz} refers to a transition of the field over
the whole universe, it has long been understood that this cannot be
the case, and that the HM solution must really correspond to a
transition over a region of roughly horizon size.  Our formalism makes
this precise, in that it is explicitly restricted to precisely a
horizon volume.  The HM solution is a $\tau$-independent solution that
happens also to be spatially constant over the horizon volume (we will
shortly encounter $\tau$-independent solutions that are not spatially
homogeneous), and so corresponds to a thermal fluctuation\footnote{ In
the thermal approach to the problem that we are taking, the stochastic
interpretation of the HM
bounce~\cite{Starobinsky:1986fx,Goncharov:1987ir,Linde:1991sk}
may be viewed as one mechanism by which a thermal distribution of
configurations is established.}  to the top of the barrier in
$V(\phi)$ (or, more properly, to a saddle point of the barrier in the
potential energy $U[\phi({\bf x})]$).  It is a curved-space analogue
of the bounce illustrated in Fig.~\ref{hiTphi}b.

\item{\bf Remote Hawking-Moss bounces:} If the potential has more than
one local maximum, then there will be HM bounce solutions corresponding
to each.  The standard expression for the transition rate, which is
unchanged in our approach, depends only on the values of $V(\phi)$ at
the false vacuum and at the local maximum, and not on the separation
in field space between the two.  If this were correct, then in a
system with many such local maxima (e.g., a string
landscape with an exponentially large number of vacua) the lifetime of
any de Sitter vacuum could be quite short. In
Ref.~\cite{Weinberg:2006pc} it was argued, from a path integral point
of view, that these bounces do not contribute.  The same conclusion
can be reached from the WKB approach we are using here, since it is
clear that the only $\tau$-independent bounces that are relevant are
those corresponding to saddle points in the barrier that immediately
surrounds the initial metastable state.

\item{\bf Oscillating bounces:} For some choices of potential the CDL
bounce equations admit solutions in which the field oscillates back
and forth across the top of the barrier in $V(\phi)$ as one moves from
one pole of the four-sphere to the other.  Examined more closely,
these are typically found to have two regions of approximate vacuum
(either true or false), one about each of the poles, separated by a
region in which the field oscillates with a relatively small amplitude
about the top of the barrier~\cite{Banks:2002nm,Hackworth:2004xb}.  If
the initial data after tunneling were obtained from a complete slice
through the bounce, these would correspond to transitions that created
not one but two vacuum bubbles, centered on antipodal poles of
the de Sitter space and separated by a region of spatially
oscillating field.  Once it is recognized that this slice must be
divided into two parts, one for the configuration before tunneling and
one for the configuration after, a much more natural picture emerges.
The oscillating bounce corresponds to a tunneling process from an
excited horizon volume configuration containing a vacuum region
surrounded by a region with the field near the top of the barrier,
through a series of intermediate configurations with the field
everywhere near the top of the barrier, and finally to a configuration
that is qualitatively like the original one except that the
final vacuum region might (or might not) be the opposite one from the
original.  We will have a bit more to say about the contribution of
these bounces in Sec.~\ref{conc}.

\item{\bf No CDL bounce:} If $V(\phi)$ is sufficiently flat near its
maximum, it can happen~\cite{Hawking:1981fz,Jensen:1983ac} that there
is no CDL bounce at all, but only a HM solution\footnote{Note that
flatness at the maximum is a necessary, but not a sufficient condition
for the nonexistence of a CDL bounce.  For examples of flat potentials
with CDL bounces, see Refs.~\cite{Hackworth:2004xb} and
\cite{Jensen:1988zx}.}.  This simply corresponds to a situation, like
that discussed at the end of Sec.~\ref{oldstuff}, in which the rate
for thermally assisted tunneling is a monotonically increasing
function of $E$.

\item{\bf Rotated bounces:} In all of the previous cases we oriented
the bounce so that it was symmetric about the $y^4$-axis. This implied
that the configurations before and after tunneling were symmetric
about the center of the horizon volume.  However, this need not be the
case.  The only condition on the symmetry axis is that it be such that
the $\tau$-derivatives of the field vanish on the $\tau=0$ and
$\tau=\pi\Lambda$ hypersurfaces.  This only requires that this axis be
perpendicular to the $y^5$-axis.  Rotating the symmetry axis away from
the $y^4$-axis gives a bounce that is not centered within the horizon
volume of interest, a possibility that one should expect.  The
surprise comes when the symmetry axis is taken to be perpendicular to
the $y^4$-axis (e.g., along the $y^3$-axis), as shown in
Fig.~\ref{twobounce}b.  In this case, the bounce has no
$\tau$-dependence at all.  It is then analogous to the critical bubble
solution of Fig.~\ref{hiTphi}b, and corresponds to a saddle point of
the static patch potential energy.  Thus, by simply rotating the
bounce we have gone from a tunneling transition to a purely thermal
one.  In contrast with the flat spacetime case, there is no sharp
distinction between the two.

\end{enumerate}

\section{Classical evolution after transition}
\label{evolution}

An important feature of the bounce formalism is that it not only
yields a tunneling rate, but also gives initial conditions for the
classical evolution after tunneling.  In addition, via a rotation from
Euclidean to Lorentzian spacetime (supplemented in some regions by
analytic continuation), it yields an actual solution of the Lorentzian
field equations.

In flat spacetime the relation between the Euclidean and Lorentzian
solutions is relatively simple.  Let us describe these by Cartesian
coordinates $(x,y,z,\tau)$ and $(x,y,z,t)$, respectively.  Now suppose
that one is given a solution $\phi_E({\bf x}, \tau)$ of the Euclidean
field equation with the property that $\partial \phi_E /\partial \tau
= 0$ everywhere on the hypersurface $\tau=0$.  One can then define initial
data for the Lorentzian equation by taking $\phi_L({\bf x},0)=
\phi_E({\bf x},0)$ and $\partial \phi_L /\partial t = 0$ everywhere on
the hypersurface $t=0$.  

Comparing the Euclidean and Lorentzian field equations, one
immediately sees that $\phi_L({\bf x},t)$ is the analytic continuation
of $\phi_E({\bf x}, \tau)$, with $t = i \tau$.  Since one does not
usually have a closed-form expression for the Euclidean solution, the
actual implementation of this continuation over all of the Lorentzian
spacetime will, in general, require explicit solution of the field
equations.  However, there may be regions of spacetime where the
analytic continuation implies equalities between the Euclidean and
Lorentzian solutions. For example, if the Euclidean solution is a
function only of $s_E=\sqrt{{\bf x}^2 + \tau^2}$, then the Lorentzian
solution is a function only of $s_L=\sqrt{{\bf x}^2 -t^2}$, and
its values in the part of spacetime lying outside the light cone of
the origin can be read off directly from the Euclidean solution.

It should be stressed that
the above are merely mathematical statements about solutions
of differential equations.  Their physical relevance depends on an 
additional fact, namely that $\phi_L({\bf x},0) = \phi_E({\bf x}, 0)$
actually is the configuration from which the classical post-tunneling
system evolves.

There are some complicating factors when we generalize this procedure
to de Sitter spacetime.  First, in contrast with the flat Minkowski
case, where the Euclidean space and Lorentzian spacetime are both topologically $R^4$, the Euclidean space and Lorentzian spacetime now have different topologies.  Second, and more importantly, there are a number of ``natural'' ways of putting coordinates on the Euclidean
space and then continuing them to the Lorentzian spacetime.   

To see the first, recall that de Sitter spacetime may
be defined as the hyperboloid
\begin{equation}
      \Lambda^2 = (y^1)^2 +(y^2)^2 + (y^3)^2+ (y^4)^2 - (y^5)^2
\end{equation}
in a five-dimensional spacetime with Minkowskian metric
\begin{equation}
    ds^2 = (dy^1)^2 +(dy^2)^2 + (dy^3)^2+ (dy^4)^2 - (dy^5)^2  \, .
\end{equation}
Its natural Euclidean counterpart is the four-sphere of radius $\Lambda$ in five-dimensional Euclidean space.

Of the various coordinate systems that can be put on the four-sphere,
two are of particular relevance to us.
The first is the generalized Hopf coordinates, defined by Eq.~(\ref{hopfdef}), in
terms of which the metric on the four-sphere is given by the
expression in Eq.~(\ref{hopfmetric}) or, if we define $\eta =
\sin^{-1}(r/\Lambda)$, by
\begin{equation}
    ds^2 = \cos^2\eta \, d\tau^2 + \Lambda^2 \left[ d\eta^2 + \sin^2\eta 
     \left(d\theta^2 + \sin^2\theta \, d\phi^2 \right) \right]  \, .
\end{equation}
These coordinates give an $S^2 \times S^1$ foliation of the
four-sphere. Making the replacement $\tau \rightarrow it$ gives
\begin{equation}
    ds^2 = -\cos^2\eta \, dt^2 + \Lambda^2 \left[ d\eta^2 + \sin^2\eta
       \left(d\theta^2 + \sin^2\theta \, d\phi^2 \right) \right] \, ,
\end{equation}
which is equivalent to the metric of Eq.~(\ref{staticCoord}).  With coordinates
restricted to real values, this covers only a portion of de Sitter spacetime, 
the causal diamond of the point $y^1=y^2=y^3=y^5=0$, $y^4=\Lambda$.

The second is the hyperspherical coordinates defined by
\begin{eqnarray}
   y^1 &=& \Lambda  \sin\xi \cos \tau' \sin\theta \cos\phi  \cr
   y^2 &=& \Lambda \sin\xi \cos \tau' \sin\theta \sin\phi
 \cr 
   y^3 &=& \Lambda\sin\xi \cos \tau' \cos\theta
 \cr \
   y^4 &=& \Lambda\cos\xi \cr
   y^5 &=& \Lambda\sin\xi \sin\tau' \, .
\label{hyperdef}
\end{eqnarray}
The corresponding metric is 
\begin{equation}
    ds^2 = \Lambda^2 d\xi^2 +\Lambda^2 \sin^2\xi \left[ d{\tau'}^2 + \cos^2 \tau' 
   \left(d\theta^2 + \sin^2\theta \,d\phi^2 \right) \right]  \, .
\end{equation} 
These coordinates foliate the four-sphere with three-spheres.
The replacement $\tau' \rightarrow i t'$ takes this to the Lorentzian
metric
\begin{equation}
    ds^2 = \Lambda^2 d\xi^2 +\Lambda^2 \sin^2\xi \left[- {dt'}^2 + \cosh^2 t'
    \left(d\theta^2 + \sin^2\theta \,d\phi^2 \right) \right]  \, .
\end{equation}
Again, real values of the coordinates cover only a portion of the de
Sitter spacetime,\footnote{ Although not especially relevant for our purposes,
there are coordinate systems on the four-sphere
that cover the full de Sitter spacetime after the rotation 
to Lorentzian time.  One such is obtained from the hyperspherical coordinates
above by interchanging $y^4$ and $y^5$, shifting $\xi$ to $\xi+\pi/2$,
and then interpreting $\xi$, and not $\tau'$, as the Euclidean time.}
in this case the region where $|y^4| \le \Lambda$.
These are the coordinates used by CDL.  They are 
are particularly convenient for studying
O(4)-symmetric bounces, because (with appropriate orientation of
axes) this symmetry is equivalent to the statement that the fields
depend only on $\xi$.

These two sets of coordinates define different hypersurfaces of
constant Euclidean time.  For the former, these are half three-spheres
(or, equivalently, three-dimensional balls bounded by two-spheres).
For the latter, these hypersurfaces are full three-spheres.
Visualized in two fewer dimensions, the former are semicircles of
fixed longitude, while the latter are circles of fixed latitude, with
the defining poles rotated $90^\circ$ in going from one case to the
other.  The constant $\tau$ and constant $\tau'$ hypersurfaces do not
generally coincide.  The one exception is the hypersurface $\tau'=0$
which, as was noted previously, is the union of the hypersurfaces
$\tau=0$ and $\tau=\pi\Lambda$.

In the CDL bounce, $\partial \phi/\partial \tau'$ vanishes everywhere
on the $\tau'=0$ hypersurface.  Mapping this hypersurface onto the
three-sphere $t'=0$ in de Sitter spacetime and continuing this
solution via $\tau' \rightarrow it'$ certainly gives a solution of the
Lorentzian field equations.  However, this solution is not the
physically relevant one, because the configuration on emerging from
tunneling is specified by the $\tau=0$ slice of the bounce.  This maps
onto the de Sitter $t=0$ hypersurface, which is only half of the
$t'=0$ hypersurface, namely the portion lying within a horizon
radius of the point $P$ with coordinates $y^1=y^2=y^3=y^5=0$, $y^4=\Lambda$.
The data outside the horizon, on the remainder of the hypersurface, is
not specified.  Hence, the future evolution of $\phi$ cannot be
determined everywhere, but only in the causal diamond of $P$, which
happens to be precisely the region covered by the continuation of the
Hopf coordinates.  It is only this restriction of the continued CDL
solution that is physically meaningful.

Because the field on the $\tau = \pi \Lambda$ slice is just the
continuation of the bounce solution from the $\tau=0$ slice, it might seem that by
keeping only the latter we are needlessly discarding initial data on
half of the de Sitter space.  This is not so.  Considered in
isolation, the data on any spacelike hypersurface are completely
unconstrained.  The fact that the data on one part of the hypersurface
are related to an interesting Euclidean solution implies nothing at
all about the data on the remainder of the hypersurface.  As our
derivation of the CDL decay rate makes clear, the bounce solution only
gives information about the fields within the static patch.  Outside this
patch, {\it any} data are possible.

When written in terms of the hyperspherical coordinates of
Eq.~(\ref{hyperdef}), O(4)-symmetric bounces that lie at the center of the causal patch (i.e., that are
     centered about the point P) are functions solely of $\xi$, and thus are determined (in any coordinate
system) solely by the value of $y^4$.  For treating bounces that are not centered about the point P, it is more convenient to use a rotated set of hyperspherical coordinates, such as those defined by
\begin{eqnarray}
   y^1 &=& \Lambda \sin\tilde \xi \cos \tilde \tau' \sin\tilde \theta \cos\tilde \phi  \cr
   y^2 &=& \Lambda \sin\tilde \xi \cos \tilde \tau' \sin\tilde \theta \sin\tilde \phi
 \cr
   y^3 &=&  \Lambda \left(\cos\alpha \, \sin\tilde \xi \cos \tilde \tau' \cos\tilde \theta
            - \sin \alpha \,  \cos\tilde \xi \right) \cr 
   y^4 &=& \Lambda \left(\sin \alpha \, \sin\tilde \xi \cos \tilde \tau' \cos\tilde \theta
            + \cos\alpha \,  \cos\tilde \xi \right) \cr
   y^5 &=& \Lambda\sin\tilde \xi \sin\tilde \tau'
\end{eqnarray}
(with $\alpha$ some fixed angle), so that the field still depends only
on the value of a single variable, in this case $\tilde \xi$. In
particular, the choice $\alpha = \pi/2$ corresponds to the
$\tau$-independent bounces discussed at the end of
Sec.~\ref{variousBounces}, as can be confirmed by comparison with
Eq.~(\ref{hopfdef}).  These depend only on $y^3$, and so after rotation to
Lorentzian space are independent of $t$.  In other words, they are
static (but unstable) solutions when viewed in static de Sitter
coordinates, in agreement with their interpretation as saddle points
of the potential energy.  In other coordinate systems, on the other
hand, these solutions are time-dependent.

\section{Concluding remarks}
\label{conc}

In this paper we have provided the question for CDL's answer.  We have
shown that their prescription for calculating transition rates between
de Sitter vacua can be obtained by considering only the field degrees
of freedom lying within the horizon-sized static patch.  Indeed,
because no reference is made to data outside the causal diamond, it is
not necessary to assume that spacetime is globally de Sitter (which it
certainly is not) or, in fact, to make any assumption about conditions
beyond the horizon.

Our approach clarifies the meaning of the bounce solution itself.  
We see that, completely in parallel with the flat spacetime case, the 
bounce gives a sequence of spatial slices interpolating between
two configurations that are turning points on opposite sides
of the potential energy barrier.
As a result, although we recover precisely the CDL result for the
transition rate, we differ from them, and from other previous
treatments (including Ref.~\cite{Guth:1982pn}), concerning the
extraction of the post-tunneling initial conditions from the bounce.
As we have seen, these are specified only within the static patch, and
not over an entire spacelike hypersurface of de Sitter space.  In
particular, our approach gives a crisp and unambiguous explanation of
why the HM solution only refers to a horizon volume.

The thermal context of our derivation shows that the CDL bounce should
be understood as a process of thermally-assisted tunneling in which
the tunneling takes place, not from the ground state itself, but from
a thermally excited state.  This explains both why
$\phi$ never achieves its false vacuum value on the bounce, and why
the bounce solution and its action are independent of the details of
the $V(\phi)$ near $\phi_{\rm fv}$.  It also leads to a natural
understanding of why some potentials admit no CDL bounces at all.

Our results also clarify the meaning of the oscillating bounce
solutions which, under the previous interpretation, would seem to
correspond to the emergence of a configuration with two vacuum
bubbles.  We now see that these bounces actually specify a path though
configuration space that connects two thermally excited horizon volume
configurations, each of which contains a single vacuum bubble.  But,
at the same time, we now have a clear argument, based on the counting
of negative eigenvalues, for discounting these.  For the flat
spacetime problem, the path integral approach depends on the fact that
the fluctuations about the bounce include one mode with a negative
eigenvalue, to provide the factor of $i$ needed to give the energy of
the false vacuum an imaginary part.  Although the factors of $i$ would
also come out correctly if there were $4n+1$ negative eigenvalues, the
WKB approach shows that bounces with more than one negative mode must
be discarded because they correspond to tunneling paths that are only
saddle points for the tunneling exponent $B$; i.e., there is a linear
combination of these modes that gives a continuous variation of the
tunneling path that lowers\footnote{This is the essence of the
argument demonstrating~\cite{Coleman:1987rm} that in flat spacetime
the bounce of lowest action has precisely one negative eigenvalue, a
result that has been extended to case with
gravity~\cite{Tanaka:1999pj,Gratton:2000fj,Khvedelidze:2000cp,Lavrelashvili:1999sr}.}
$B$.  Now that we have an extension of the tunneling path approach to
the curved spacetime problem, we see unambiguously that bounces with
multiple negative eigenvalues should not be included.  Since
oscillating bounces always have multiple negative
eigenvalues~\cite{HackworthThesis,Lavrelashvili:2006cv}, they must be
excluded.

We can also see that similar remarks apply to the HM solution when it
has more than one negative eigenvalue [i.e., when $V''(\phi) > 4
\Lambda^{-2}$ at the top of the barrier].  When it has only one
negative eigenvalue, the HM solution corresponds to a local minimum in
the barrier surrounding the false vacuum.  Any additional negative
eigenvalues denote the existence of directions in which the barrier
decreases, thus indicating that there must be some lower saddle point
along the top of the barrier. This saddle point is a spatially
inhomogeneous static solution corresponding to a rotated bounce of the
sort described in Sec.~\ref{variousBounces}. This explains why the
appearance of multiple negative modes about the HM bounce is always
accompanied by the appearance of a new CDL
solution~\cite{Hackworth:2004xb}.  (Note, however, that the existence
of a CDL solution does not necessarily imply that the HM solution has
multiple negative eigenvalues.  There are potentials for which a HM
solution with just one negative eigenvalue coexists with a CDL bounce,
so that there are two competing modes for completing the transition.)
 
Our approach can be applied to transitions in anti-de Sitter
spacetime, again under the assumption that the variation in the
potential between the two vacua is small compared to its absolute
value.  Because there is no horizon, the tunneling problem in the
fixed background limit is analogous to that in flat spacetime, with
the Euclidean time taking on all real values and no periodicity
condition.  As is well-known, for certain choices of potentials there
is no bounce solution connecting the true and false
vacua~\cite{Coleman:1980aw,Parke:1982pm}.  This can be understood by
recalling that at zero temperature the tunneling path must connect the
initial vacuum with a configuration of equal energy.  In flat
spacetime this is always possible, because a bubble of true vacuum can
always be made large enough that the energy gain from converting false
vacuum to true compensates for the energy in the bubble wall.  This is
not always so in anti-de Sitter space, because for radius much greater
than the curvature length the volume and area grow at the same rate.
Although there is a spatially homogeneous configuration on the true vacuum side
of the barrier that has the same energy as the homogeneous false
vacuum, it is inaccessible because the two are separated by an
infinite potential energy barrier, just as are degenerate field theory
vacua in flat spacetime.

Clearly the primary issue to be addressed is the extension of the
method to the more general case, where the approximation of a fixed
background geometry is not applicable.  Doing this will require the
resolution of some significant technical issues.  However, we have
seen that the problematic aspects of the CDL formalism that were
described in the introduction can be understood even when the
gravitational background is taken to be fixed, and we expect these
qualitative features to persist when the gravitational degrees of
freedom are included.  In particular, the configuration after
tunneling should still be determined only on a causal region, and
should still be obtained from a partial, not a full, slice through the
CDL bounce.  We hope to return to this in a future publication.

\begin{acknowledgments}

We are grateful for illuminating conversations with Puneet Batra,
Matthew Kleban, Kimyeong Lee, Alex Maloney, and Piljin Yi.  EJW thanks
the Institute for Advanced Study, where part of this work was done,
for its hospitality. ARB was supported by a NSF Graduate Research
Fellowship.  This work was supported in part by the Monell Foundation,
the National Science Foundation under grant PHY-0503584, and the US
Department of Energy.

\end{acknowledgments}

\end{document}